\documentclass[runningheads]{llncs}
\usepackage{graphicx}
\usepackage{array}

\begin{document}
%
\title{Auction Type Resolution on Smart Derivatives
}
%
%
\author{Yusuke Ikeno\inst{1} \and
 James Angel \inst{1} \and
 Shin'ichiro Matsuo\inst{1} \and
 Ryosuke Ushida\inst{1}}

\authorrunning{Y. Ikeno et al.}%
%
\institute{Georgetown University, \\
\email{\{yi106,angelj,sm3377,ru64\}@georgetown.edu}\\}
\maketitle              
\begin{abstract}
This paper proposes an auction type resolution for smart derivatives. It has been discussed to migrate derivatives contracts to smart contracts (smart derivatives).  Automation is often discussed in this context. It is also important to prepare to avoid disputes from practical perspectives.  There are controversial issues to terminate the relationship at defaults.In OTC derivative markets, master agreements define a basic policy for the liquidation process but there happened some disputes over these processes. We propose to define an auction type resolution in smart derivatives, which each participant would find beneficial. 

\end{abstract}
\section{Introduction}
\subsection{Background}
OTC derivatives are traded in the financial market for the purpose of hedging or transforming market risk and speculations. According to BIS statistics, notional amounts outstanding in the first half of 2020 is 606,810 billion US dollars \footnote{\url{https://www.bis.org/statistics/derstats.htm?m=6\%7C32\%7C71}}. 
This figure is significant since the global GDP in 2019 is 87,799 billion (the World Bank data\footnote{\url{https://data.worldbank.org/indicator/ny.gdp.mktp.cd}}). 

After the financial crisis, the regulation for OTC derivatives had got harder to stabilize financial markets. This made the post trading process more complicated.
Not only financial institutions  but also non financial derivatives users are suffering from the regulation cost.

The financial industry finds that distributed ledger technology and smart contracts could improve their business. The purpose is not only to make their operations more efficient but also pioneer new area.
For example, JPMorgan developed "Quorum" and Barclays held some hackasons called "DerivHack". There are many research cases globally.
Recently many central banks are studying digital currencies, which would have a significant impact on OTC derivatives as one of the use cases.\cite{CPMI}

Though there are many types of products, it is well known that they are composites of conditional cash flows or settlements.There are some studies that show they can be expressed in a functional programming framework.\cite{frankau_spinellis_nassuphis_burgard_2009} This feature also inspires it is appropriate to bring these products into smart contracts.

\subsection{Related works}
Morini discussed re-design of the buisiness model noting that "Blockchain technology was created to change some trust-based business processes to make them less reliant on trust". \cite{RePEc:ris:jofitr:1587} 
He also investigated the implementation of collateral settlement  on public blockchain. \cite{Morini2}

ISDA had published a series of guidelines for smart derivative contracts from legal perspectives .\cite{ISDAIntro}

Fries and Kohl-Landgraf thought regulatory implementations are inefficient due to  historic infrastructures and lack of standardization and automation of a trade life cycle processes.\cite{Fries3} 
They also implemented prototypes under some assumptions of digital currencies. \cite{2019arXiv190300067F,Fries2}

Clack and McGonalge discussed how smart contract code might process and automate payments-related and deliveries-related events and proposed  we should understand derivatives contracts at different levels, in terms of both the legal documentation and the work-flow. \cite{2019arXiv190401461C}  

\subsection{Contributions}
It is important to process close-out effectively because events of defaults affect significantly.  Though master agreements define basic processes to close out, calculation methodology is vague to some extent. 

Clack and Datoo recently discussed  how we can make smart close-out efficiently from legal perspectives. \cite{2020arXiv201107379D}
We discussed this issue from operational viewpoint.
As the liquidation process at default contains conflicts between involve parties by nature, the stakeholders are likely to take actions to protect their profit. 
The current method investigates the mid point under possibly transparent manner.
We proposed an auction method to calculate the termination amounts and look for new trades so that the involved parties can process more algorithmically and each participant can find merits.

\section{Practice in OTC derivatives }
\subsection{Model of OTC derivatives over blockchain}
Our discussion focuses on uncleared collateralized OTC derivatives.  
When two parties start to trade OTC derivatives, they usually sign a "master agreement", which governs all derivatives contracts between them. 
ISDA (International Swap and Derivatives Association) provides templates for the master agreement.  
Master agreements treat governed individual transactions as one contract, which make netting settlements possible. 
The definition of jurisdiction and basic policies for resolving disputes are also described in master agreements.
Schedule and credit support anex (so called "CSA") are the related documents for precise information and the rule for collateral management, which complement the master agreement.

\begin{figure}
 \centering
 \includegraphics[keepaspectratio, scale=0.75]
      {"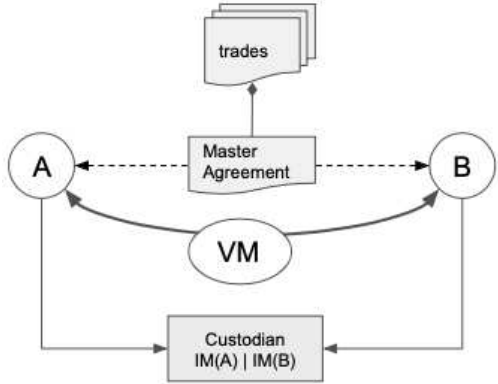"}
 \caption{Trading scheme}
 \label{fig1}
\end{figure}

From here, we review the basic process after events of default. See \cite{ISDA} for more details.
Master agreements define events of default as below. 

\begin{itemize}
    \item[$\diamondsuit$] Events of Defaults
    \item Failure to pay or deliver
    \item Breach/ repudiation of agreement
    \item Credit support default
    \item Misrepresentation
    \item Default under specified Transaction
    \item Cross-default
    \item Bankruptcy
    \item Merger without Assumption
\end{itemize}

Generally speaking, an event of default occurs if one of the parties is at fault.

If an event of default occurs (the party at fault is called "defaulting party"), the other party called "non defaulting party" has a right to terminate the contracts under their master agreement. The legal enforceability of this right depends on the jurisdiction. The party usually request legal opnions. This point is discussed in \cite{2020arXiv201107379D}.

When the right is exercised, both parties enter into "Early Termination" as the agreement describes.

\begin{enumerate}
    \item Designation of early termination date
    \item Cessation of payment and delivery obligations
    \item Occurence of early termination
    \item Calculation of early termination amounts
    \item Delivery of statement detailing net termination amount payable
    \item Payment of termination amount
\end{enumerate}

As described precisely below, the termination amount is the summation of defined payments and present values of defaulted transactions. Since someone has to calculate the present value, it is usually expected that the non defaulting party asks "Reference Market Makers" to quote from neutral positions.

If an event of default occurs (and the right is exercised), the non defaulting party loses its risk exposure coming from those defaulted transactions.   The non defaulting party will cover the exposure in the market.   If there are only a few defaulted transactions, it is easy to cover at once.   Otherwise it takes time to reconstruct the exposure as there might be concerns on market impact and liquidity.   This is why initial margin is required.   IM is theoretically designed to absorb such timing and liquidity risk.   But we should understand that the cost might exceed the margin depending on market conditions even though the model explains we rarely face excess loss.

\begin{figure}
 \centering
 \includegraphics[keepaspectratio, scale=0.8]
      {"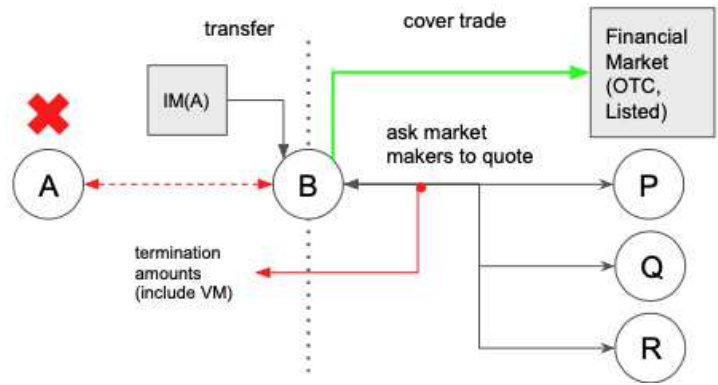"}
 \caption{Covering the defaulted exposures}
 \label{fig2}
\end{figure}

Especially for early termination, the parties should
\begin{itemize}
    \item terminate transactions under the agreed value.
    \item close the process as soon as possible
    \item reduce the effect to the financial system
\end{itemize}

The master agreements give a policy for these points but the exact way is up to the involved parties. There could be conflicts at termination amounts between both parties because the profit of one party means the loss of the other.  Unfortunately there were some cases the defaulting party started  disputes even though the quoted amount should have been accepted according to the agreement. 

Since the market circumstance might be stressed after events of default, it is helpful for stakeholders to reconsider the process depending on the situation.
We should investigate the way satisfying
\begin{enumerate}
    \item the predefined process will be executed algorithmically within expectation.
    \item the process can be paused if unexpected loss would happen.
    \item each participant can cooperate in a constructive manner.
\end{enumerate}

\subsection{Credit risk in uncleared trades}
\subsubsection{Mechanism}
The price movement in financial markets affects the valuation of contracts like other financial products.  Unrealized profit (or loss) would be lost if the counterparty fails to continue.   In other words, derivative contracts include counterparty's credit risk.

If markets move significantly, we should take care of not only our profit or loss but also how counterparties are working.   

\subsubsection{Margin}
New York Fed describes the history of the regulation of derivatives as below; 
In 2009, the G20 Leaders agreed to reforms in the OTC derivatives market to achieve central clearing and, where appropriate, exchange or electronic trading of standardized OTC derivatives; reporting of all transactions to trade repositories; and higher capital as well as margin requirements for non-centrally cleared transactions \footnote{New York Fed  "Over-The-Counter Derivatives"https://www.newyorkfed.org/financial-services-and-infrastructure/financial-market-infrastructure-and-reform/over-the-counter-derivatives"}.  The margin requirement on uncleared OTC derivatives is yet on the way.

There are two types of margin.   One is called "variable margin"(VM), which covers present values of contracts.  If your position has profit, you will receive some assets as collateral whose value matches your profit. If loss, you will have to transfer to your counterparty. 

The other one is called "initial margin"(IM), which is the cost for reconstructing the positions.  Under the margin requirement for uncleared OTC derivatives, IM are required to be deposited into segregated  accounts from each party, such as custodians.   We need statistical models to calculate this amount.   The SIMM, which is now regarded as a standard margin model for uncleared OTC derivatives, is designed to "meet a 99\% confidence level of cover over a 10-day standard margin period of risk". \cite{Rama}

\subsection{Issues for calculation of termination amount}
The calculation includes  defined unpaid amounts that were due before the early termination date and present values of contracts at the early termination date.   The early termination amount is usually calculated by the non defaulting party.    (There are some other cases which both parties agreed.    

The "Market Quotation" is used for calculation of the amount to protect each profit .   The master agreement defines that “Market Quotation”  is an amount determined on the basis of quotations from Reference Market-makers.   "Reference Market Makers" is also defined as following; “Reference Market-makers” means four leading dealers in the relevant market selected by the party determining a Market Quotation in good faith (a) from among dealers of the highest credit standing which satisfy all the criteria that such party applies generally at the time in deciding whether to offer or to make an extension of credit and (b) to the extent practicable, from among such dealers having an office in the same city. \cite{masteragreementsample}   In short, "Reference Market Makers" are those trusted third parties who can provide fair calculation of derivative transactions.   We can imagine which companies are suitable for this role but it is uncertain that those companies are willing to join the liquidation process.

Even though the process does not permit claims from default parties, there were some cases that a defaulting party had disagreed with the amount, which caused long lasting disputes. Since the defaulting party can be motivated to raise their value at the events of default, it is effective to provide the party another incentive so that they can close smoothly. 

\subsection{CCP cases}
In case of centrally cleared trades, the derivatives transactions of defaulted members could be transferred to other clearing members via the auction process in a few days.

For example, LCH, one of the largest clearing house, maintains the default process as below\footnote{https://www.lch.com/services/swapclear/risk-management
};
\begin{enumerate}
    \item[(1)] Risk neutralisation \& client porting: \mbox{}\\
    After a Clearing Member default, LCH immediately begins porting non-defaulting clients to non-defaulting Clearing Members. Alongside this, we hedge the defaulter’s trade portfolio with the assistance of our Default Management Group (DMG), senior executives with appropriate skills and expertises from its clearing members and certain other members as the DMG considers appropriate, that are seconded to LCH in the event of default.
    \item[(2)] Portfolio auction: \mbox{}\\
    Once the risk of the portfolio is substantially reduced by the DMG, LCH splits the defaulter’s portfolio by product and currency. An auction is then conducted for each portfolio. The ability to receive and price an auctioned portfolio is one of the criteria we verify prior to granting firms membership to SwapClear.
    \item[(3)] Loss attribution: \mbox{} \\
    In the event that losses are greater than the financial resources of the defaulting member and of LCH, the funded Default Fund contributions of non-defaulting SwapClear members are used.
\end{enumerate}

\section{Auction resolution}
There are several types of auctions implemented on the smart contracts.\cite{10.1007/978-3-030-16946-6_5,conf/fc/GalalY18} 
The main purpose in this case is to investigate the fair mid value of transactions and liquidity in the market. 
We use the same algorithm for the mid value as the "Market Quotation". 
A second price sealed bid auction is applied to choose the winner of trade value. 
We require a stopping condition about the trade cost and IM, which each participant finds beneficial to cooperate.

\begin{figure}
 \centering
 \includegraphics[keepaspectratio, scale=0.8]
      {"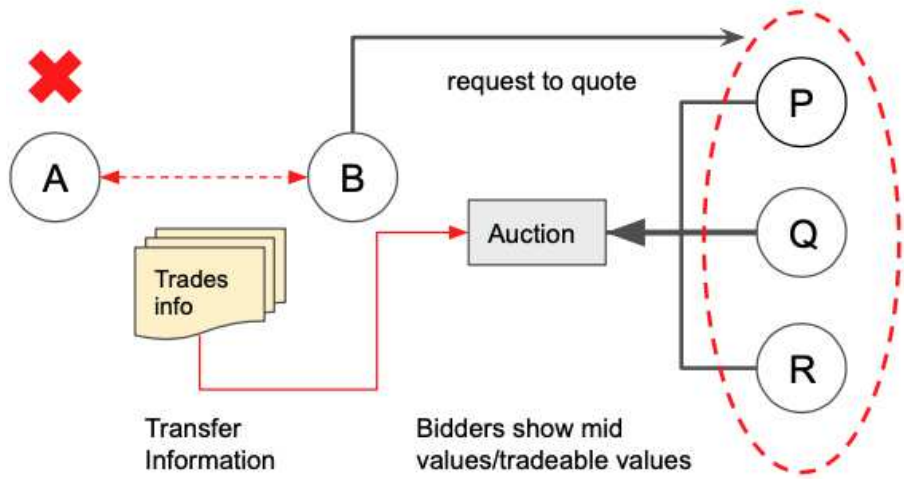"}
 \caption{Initiate an auction}
 \label{fig3}
\end{figure}

\subsection{Workflow}
Based on the current workflow, we would like to open an auction to investigate the fair values for termination amounts.
\begin{itemize}
    \item When the non defaulting party exercises the right, the auction contract is executed.
    \item The non defaulting party asks tradable entities from its relationship to join the auction,
    \item Bidders are required to show the mid valuation. If possible to trade, they can also show the tradable value.
    \item After bidding, the smart contract calculates the mid valuation (=MQ) in the predefined manner.   (For example, averaging excluding outliers for the mid values.) 
    \item The defaulted trades would be terminated at MQ.
    \item As for the trade value, the second highest bid is chosen.(second-price sealed-bid auctions) 
    \item If the trade cost ($:= | \mathrm{the \ mid \ value} - \mathrm{the \ trade \  value}|$) is more than IM, the non defaulting party has a right to avoid to trade.
        \begin{itemize}
        \item As described before, IM is supposed to compensate for the potential loss, which can be understood as a cost to reconstruct the exposure.
        \item In this auction, the difference between the mid value and the trade costs is the cost that the non defaulting party has to pay for.
        \item If the cost is more than IM, the non defaulting party might find it advantageous to cover in the market directly.
        \item If the cost is less then IM, the residual can be reverted to the defaulting party, which could incentivises the defaulting party to agree with this auction process.
    \end{itemize}
\end{itemize}

Since the optimal strategy in second bid auctions is showing the maximum value which they are willing to pay, the non defaulting party can redeem the realistic value for package trades.


It looks curious to revert the residual to the defaulting party because IM is supposed to transfer to the non defaulting party by nature. As described before IM is the potential loss to cover exposures arising from defaulted trades. 
If the new trades are given without costs, the non defaulting party does not require this margin for that purposes.
We can think it to revert the residual to the defaulting party if the termination amounts are payable.



\begin{figure}
 \centering
 \includegraphics[keepaspectratio, scale=0.65]
      {"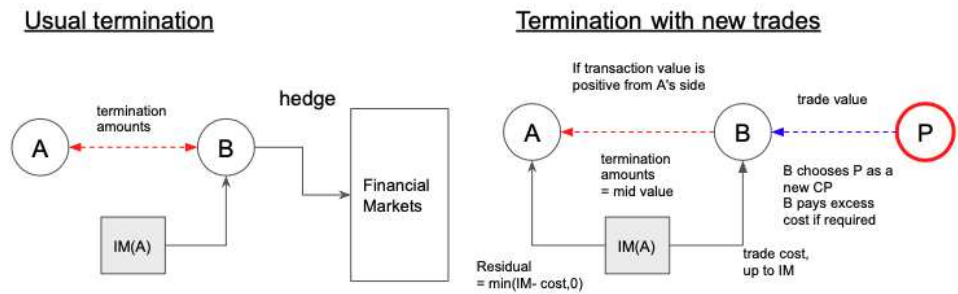"}
 \caption{Settlement of termination amounts}
 \label{fig4}
\end{figure}

The following Table \ref{tab1} shows some sample cases. Each value represents the transactions value from A (the defaulting party). There are four or five bidders depending on the cases. The "mid" value is quoted for the "Market Quotation" and the "trade" value is the price at which the bidder is willing to trade. The "aggregate" value means the "Market Quotation" for mid values and the second price auction winner and the price for trade values, leaving IM amount aside.


\begin{table}
\caption{Sample auction scenarios.}\label{tab1}
\begin{tabular}{|r|r|r|r|r|r|r|r|r|r|}
\hline 
           &  &  &      & \multicolumn{5}{|l|}{bidders} &  \\ \cline{5-9} 
No    & A & B & \ \ \  value   & 1    & 2 & 3 & 4 & 5 & \ \ aggregate \\ \cline{1-10}
1           &   \ \ \ \ 100 & \ \ \ \ 80 & \ \  mid & \ \ \ \ 90.00   & \ \ \ \ 95.00&  \ \ \ \ 85.00 &  \ \ \ \ 100.00 &  \ \ \ \ 90.00 &  \ \ \ \ 91.67 \\ \cline{4-10}
 & & & trade & 85.00 & - & 80.00 & 90.00 & - & \ \  ($\sharp$ 4, 85.00) \\ \cline{1-10}
2           &   \ \ \ \ 100 & \ \ \ \ 80 & \ \  mid & \ \ \ \ 90.00   & \ \ \ \ 95.00&  \ \ \ \ 85.00 &  \ \ \ \ 100.00 &  \ \ \ \ - &  \ \ \ \ 92.50 \\ \cline{4-10}
 & & & trade & - & - & 80.00 & - & - & ( $\sharp$ 3, 80.00) \\ \cline{1-10}
3           &   \ \ \ \ 100 & \ \ \ \ 80 & \ \  mid & \ \ \ \ 95.00   & \ \ \ \ 95.00&  \ \ \ \ 75.00 &  \ \ \ \ 105.00 &  \ \ \ \ - &  \ \ \ \ 95.00 \\ \cline{4-10}
 & & & trade & 90.00 & - & - & 95.00  & - & ($\sharp$ 4, 90.00) \\ \cline{1-10}
4           &   \ \ \ \ 100 & \ \ \ \ 80 & \ \  mid & \ \ \ \ 90.00   & \ \ \ \ 90.00&  \ \ \ \ 100.00 &  \ \ \ \ 100.00 &  \ \ \ \ 90.00 &  \ \ \ \ 93.33 \\ \cline{4-10}
 & & & trade & - & 75.00 & 80.00 & -  & - & ($\sharp$ 3, 75.00) \\ \cline{1-10}
 5          &   \ \ \ \ 100 & \ \ \ \ 80 & \ \  mid & \ \ \ \ 90.00   & \ \ \ \ 90.00&  \ \ \ \ 90.00 &  \ \ \ \ 90.00 &  \ \ \ \ 90.00 &  \ \ \ \ 90.00 \\ \cline{4-10}
 & & & trade & - & - & 88.00 & -  & - & ($\sharp$ 3, 88.00)  \\ \cline{1-10}
\end{tabular}
\end{table}

Depending on the market liquidity, B might want immediate counterparties to cover its exposure. In such a case, B could ask the bidder after the process because the cancel is a result of algorithm. Though there could be some critics that this is not fair for the defaulting party, IM covers these covering risk by its definition. The residual is a bonus for the defaulting party.

\subsection{Requirements}
\subsubsection{Standardization of information}
Since bidders have to calculate the value of transactions by themselves, it is important to share the transactions and related information.
\begin{itemize}
    \item Both parties names
    \item Master agreement and related documents \mbox{} \\
    The content of master agreement and related documents depends individually. For example, collateral information affects valuation directly.
    \item Details of each transaction \mbox{}\\
    Exotic trades might be included among the defaulted trades as they are not cleared.
\end{itemize}

If we could record this information in the same manner, it would be helpful to share. There are some formats such as FIX and FpML to describe the transactions. The booking system records the transactions in its original format based on these common formats. Sharing details of transactions require some kind of transformation.

When two parties start to trade OTC derivatives, they negotiate the details of master agreement based on the template.
The amendment of template causes a potentially infinite pattern of master agreement and the customization makes the initial negotiation annoying.

Under the digitization of derivatives, ISDA develops "Common Domain Model" (so-called "CDM")  and "Clause Library".\cite{2017arXiv171110964C,ISDACDM,ISDACL}  CDM helps the users to describe the transactions in a unified manner and Clause Library will reduce the cost of legal negotiations.
 
\subsubsection{Invitation of bidders}
Bidders with good faith are essential to make these auctions successful.
We can not exclude the case of no bidder cases and manipulations by bidders.

In the current process, reference market makers are assumed to quote market quotations. We can assume they will join to quote  the mid values faithfully.
In other words, the existence of reference market makers is the trust point of this process.
If both parties could not find out such parties, they would  have to negotiate their termination amounts by themselves. They might need enforceable authority such as a court.

\section{Evaluation}
If both parties involved in OTC derivatives 
As this auction satisfies the current price decision process ( "Market Quotation"), this can be understood as the improvement. 

If the tradeable value is available, the non defaulting party can avoid liquidity concerns and the defaulting party may be able to retrieve some of the IM. 

Bidders may find it attractive to take part in this auction if they can reduce their risk exposures at lower cost. Though this depends on the market situation, their stance could send a signal of liquidity to the non defaulting party.

It is basically beneficial for each participant to open an auction as a part of the liquidatuin. 

\section{Discussions}
\subsection{Auction style}
We select the second price sealed auction for trade value to clarify the meaning of bids.
The policy is bit different from calculation of the mid value, which investigates the average value.
Those participants whose calculate relatively higher are likely chosen as the winners.
Higher price is nice as the defaulting party gains more residual and it costs less for the non defaulting party to cover the exposure at once.
The difference of the valuation could cause another problem, for example collateral process. Under the collateral exchange, two parties are required to exchange VM reflecting each exposure calculation. If the valuation is far, they meet disputes. There are some temporal resolution such as averaging or undisputed amount.\cite{simmons2019collateral}


Since CCP can not keep the defaulted trades, they have to immediately transfer the portfolio  to another member chosen by an auction. In this case, VM and IM are alos transferred to the new party at the same time. 
If we totally imitate this, the situation is very simple but we could not find out the merit from the defaulting party.
CCP scheme is strongly supported by the duty of clearing members.

OTC trades are basically closed bilaterally. 
It is difficult to define the "fair value" in OTC markets because the contract data is private by nature so that we have to rely on the market participants at both entry and exit points.
As the master agreement assumes reference market makers, it is important to manage risk in the community.

\subsection{Electronic platforms}
Electronic platforms (e.g., TradeWeb) support efficient executions, on which users can request for a quote of not only one trade but also several trades.   
Some may wonder if the non defaulting party could request for "Market Quotation" via these platforms efficiently instead of opening an auction.
What we emphasise is the predefined process to gain the fair value on the smart contracts.
Using smart contracts do not depend on a going concern principle, which could eliminate a kind of trust.
Of course  there may exist the more efficient way to combine the execution platform  with smart contracts.

\subsection{Governance of the auction process}
We assume that the process is governed by the smart contract.  Inputs and decisions are given by participants. 
We must rely on the bidders to succeed in the auction. 
As the current procedure assumes there are referenced market makers, we also assume trusted quotes here.

\subsection{Privacy}
If eligible tradeable prices do not appear in the auction at all, the non defaulting party has to trade in the market directly. 
In this case bidders know the defaulted risk exposure, which might make it difficult for the party to trade efficiently.
As we assume the non defaulting party invites the bidders in this scheme those parties are unlikely do so.
Though this is a kind of morality,  we should take some measure against such immoral players in future.

On the other hand, this information might have a kind of signaling effects, so that some faithful market makers support to close the position.

Pricing OTC derivatives can be expressed as a real-valued mapping whose domains are transactions, valuation models and market data. If there were a contract which could calculate the value from transactions (provided by the non defaulting party) and the pair of a valuation model and market data (provided by the bidders), we would not concern the privacy. From this point, standarzation of valuation matters.

\section{Conclusion and future works}
Since we discussed the concept to improve the process in this paper, we should implement this idea to investigate more practically.

For efficient valuation, it is more convenient to assume trusted valuation agents independent from dealers; in other words, oracles pricing the transactions.    If it is difficult to set up agents which calculate automatically, we may prepare for the shared calculation methodologies  in advance. Fries et al \cite{2019arXiv190300067F} also pointed out this point.


Decentralised finances (so called "DeFi") provide derivatives for crypto assets. They are usually overcollateralized and lock the potential cash flow at first to avoid any disputes. If users request more flexibility as traditional derivatives, the similar issues including dispute resolutions are required. In this area, the algorithmic methodology would be prefered.  

We assume some parties would quote the tradable value because the current process also assumes "reference market makers".  As financial markets might face less liquidity at events of defaults, the non defaulting party would not find out the candidates for its auction.  In such a case, the IM could not compensate for the cover cost. We have to continue monitoring how the situation will change in future. Even though we could define reasonable resolutions, the involved parties might change their mind at the liquidation. We would also rely on other enforceable resolutions which work with the algorithm. 
Under the context of digital transformation, smart contracts would play important roles to automate and redefine many workflows.  It is important to not only enjoy its efficiencies but also scrutinize its potential disputes and resolutions from practical viewpoints.

%
%
%
%

\bibliographystyle{splncs04}
\bibliography{WTSC_subm}

\end{document}